\documentstyle[amssymb,prl,aps,epsfig]{revtex}

\begin{document}

\title{Potential flows in a core-dipole-shell system: numerical
results}

\author{Maximiliano Ujevic \thanks{e-mail: mujevic@ime.unicamp.br}
and
Patricio S. Letelier\thanks{e-mail: letelier@ime.unicamp.br} }

\address{Departamento de Matem\'atica Aplicada, Instituto de
Matem\'atica, Estat\'{\i}stica e Computa\c{c}\~ao Cient\'{\i}fica, 
Universidade Estadual de Campinas,  13081-970, Campinas, SP, Brasil}


\maketitle

\begin{abstract}
Numerical solutions for:  the integral curves of the velocity field
(streamlines), the density contours, and the accretion rate of a
steady-state flow of an ideal fluid with  $p=K n^\gamma$ equation of
state  orbiting in  a core-dipole-shell system are presented. For
$\gamma\not = 2$, we found that the non-linear contribution appearing
in the partial differential equation for the velocity potential  has
little effect in the form of the streamlines and density contour
lines, but can be noticed in the  density values. The study of several
cases indicates that this appears to be  the general situation. The
accretion rate was found to increase when the constant $\gamma$
decreases.

\end{abstract}

\maketitle


\section{Introduction}

In a recent work \cite{uje:let} we study the streamlines and density
contours of a stationary  fluid in the presence of either a rigid
sphere or  a black hole both with a  distant  shell of matter modeled
by a dipolar field. To fix ideas see Fig. 1 where this situation is 
depicted.
  The substitution of a halo or shell of matter by 
a fixed axisymmetric field is an old and common method in astrophysics, 
 for example to simulate galaxy halos \cite{hoc:bro}. Numerical 
experiments  confirm that this is usually a good approximation \cite{hoh}.
 This method can be used in other physical situations 
where a core and a distant shell of matter are present.

In \cite{uje:let} we used
for the fluid the stiff equation of state, i.e., a polytropic,  $p=K
n^\gamma$ with $\gamma=2$.  This stiff equation of state over
simplifies the partial differential equation for the velocity
potential field $\Phi$ (it becomes linear). In this  case the sound
velocity in the fluid is equal to the speed of light. Therefore the
stiff equation of state represents a limit situation not easily
encountered in usual  physical situations.  Abrahams and
Shapiro \cite{abr:sha}  studied  the  more realistic situation,  a
fluid with polytropic equation of state with   $ 1<\gamma<2 $ in the
presence of either a rigid sphere or  a black hole.  The sound
velocity in this case is less than the speed of light and  the partial
differential equation for $\Phi$ turns to be  nonlinear.

In the present  work,  following  Abrahams and Shapiro \cite{abr:sha},
 we study  a more realistic situation  than the one examined in
 \cite{uje:let}.   Now we  compute  the streamlines and baryon density
 contours for a  fluid with polytropic
equation of state with
 $1<\gamma<2$ in the presence of  either a rigid  sphere or a  black
 hole both with dipolar halos.  In other words, now we consider the
 same situation as before, but for a  general polytropic fluid.  We
 consider also the accretion rate of particles into the black hole at
 its dependence with  $\gamma$.  Furthermore, we  complete the work of
 Abrahams and Shapiro \cite{abr:sha}  by examining  the black hole
 case with $1<\gamma <2$ and asymptotic constant velocity  (case not
 studied in the quoted work). In particular, we study  streamlines,
 density contours and accretion rate.

 In section II we present the basic equations that describe potential
flows and the nonlinear  equation for the velocity potential for a
polytropic equation of state of the form $p=K n^\gamma$. Section III
is divided in three sub-sections. In sub-section III-A we introduce
the metric that represents a central core with a distant dipolar
shell of matter. This field is the first approximation to represent
distant matter, like  halos and rings. In sub-section III-B we present
the numerical method used to solve the  nonlinear partial differential
equation presented in section II. In sub-section III-C we show some
results and study the behavior of the streamlines and density contours
of the fluid  in the presence of either  a rigid sphere or a  black
hole both with a dipolar halo. Also we compute  the accretion rate of
particles falling  into a black hole with and without dipolar
halo. Finally, in section IV we summarize our results.

\section{Basic Equations}

The solution for potential flow of  the equation of motion derived
from the energy-momentum tensor for an ideal fluid,  $T_{\mu\nu} = (p
+ \rho) U_\mu U_\nu + p g_{\mu\nu}$, is
\begin{equation}
h U_\alpha = \Phi_{,\alpha}, \label{potflow}
\end{equation}
\noindent where $h=(\rho +p)/n$ denotes the enthalpy per baryon, $n$ the 
baryon density, $p$ the pressure of the fluid, $\rho = \rho_0 + \epsilon$  
the total energy density, $\rho_0$ the rest mass energy  density,  
$\epsilon$ the internal energy density, $U^\mu$ represents the fluid  
four-velocity, and $\Phi$ the scalar potential. Our conventions are 
$G=c=1$, metric with signature +2. Partial and covariant derivatives are  
denoted by commas and semicolons, respectively.

We  recall that, by definition, an  ideal fluid is  adiabatic 
\cite{note1}, i.e. the entropy is constant along the world line of each 
fluid element.  If in an instant $t$ the entropy is constant throughout  
the volume of the fluid (isentropic case) then, for all time and any
subsequent motion of the fluid,  it  retains everywhere the same
constant value. In that case  we find the solution for potential flows
(\ref{potflow}). Hence,  only isentropic flows can have potential
flows \cite{lan:lif}.

From Eq. (\ref{potflow}) and the baryon number density  conservation
equation,
\begin{equation}
(n U^\mu)_{;\mu}=0, \label{consern}
\end{equation}
we obtain the differential equation for the velocity potential,
\begin{equation}
\Box \Phi + [\ln \left( \frac{n}{h} \right) ]_{,\alpha}
\Phi^{,\alpha}=0, \label{diffeqn}
\end{equation}
 where, $\Box \Phi = [ \sqrt{-g} g^{\mu \nu} \Phi_{,\mu}
]_{,\nu} / \sqrt{-g}$. This equation is in general nonlinear and depends 
on the fluid equation of state. If we consider a polytropic equation of 
state $p= K n^{\gamma}$ and use the first law of thermodynamics for
 isentropic flows, ${\rm 
d} (\rho/n) + p {\rm d} (1/n) = T {\rm d} s$,  
($s$ is the entropy per baryon and $T$ the temperature), we find
\begin{equation}
\rho = \rho_0 + \frac{Kn^\gamma}{\gamma-1}.
\end{equation}
Now,  assuming that the  flow is relativistic we can neglect
the  rest mass energy density $\rho_0$ with respect to the internal
energy density $K n^\gamma/(\gamma-1)$, i.e. the  flow satisfies a
barotropic   equation of state, $p=(\gamma -1)\rho$. In
this case, the baryon number density can be written in terms  of the 
enthalpy as
\begin{equation}
n= \kappa h^{{1}\over{\gamma -1}}, \label{density}
\end{equation}
\noindent where $\kappa = \left( \frac{\gamma -1}{\gamma K} \right) 
^{1/(\gamma-1)}$. Then, equation (\ref{diffeqn})  can be written in the 
form \begin{equation}
\Box \Phi + {{2 - \gamma}\over{\gamma -1}} [\ln h(\Phi)]_{,\alpha}
\Phi^{,\alpha}=0 \label{diffnlin},
\end{equation}
\noindent The simplest  case is found when $\gamma = 2$, Eq. 
(\ref{diffeqn}) reduces to a  linear equation. In this case the
barotropic sound speed, defined as ${\rm d}P/{\rm d}\rho \equiv
c_s^2$, is equal to  the speed of light, i.e. we have a stiff equation
of state.  In the  general case  ($1<\gamma <2$), the differential
equation is nonlinear and  the barotropic sound speed is $c_s^2=(\gamma
-1)$ is   less than  the speed of light.

The normalization condition, $U_\alpha U^\alpha =-1$, gives
us a relation between the enthalpy and  the scalar field,
\begin{equation}
h=\sqrt{-\Phi{,_\alpha} \Phi^{,\alpha}}. \label{enthalpy}
\end{equation}
\noindent This relation will be also useful to determine  the baryon
number density.

In this section  we have  followed the work of  Moncrief \cite{mon}
which is in accord  with  Tabensky and Taub \cite{tab:tau} that
consider a constant barotropic sound speed,  $c_s^2=(\gamma -1)$.  We
note that   Abrahams and Shapiro \cite{abr:sha}  consider a variable
polytropic sound speed.

\section{Potential flows in black holes and rigid spheres with dipolar
halos}

\subsection{The metric}

To incorporate a dipolar field in the Schwarzschild metric we consider
the solution found in \cite{vie:let} where  authors model the
intermediate vacuum between a core and a distant shell of matter. This
core-shell system was found solving the vacuum Einstein equations  for
a general static axially symmetric metric (Weyl solution \cite{syn}).
One of these equations can be solved in terms of Legendre polynomials
increasing with the distance.  These shell-like structures can be 
used to model many situations of interest in astrophysics, such as: 
supernovas, nebulae, and galaxy halos  that have a core and an
exterior shell of matter. As a first approximation we will consider
only the first term in the expansion, the dipolar term that in
Newtonian gravity corresponds to a constant force.  By letting in the
solution presented in  \cite{vie:let} the quadrupole and octopole
moments equal zero, we obtain
\begin{eqnarray}
ds^2 &=& -\left( {{u-1}\over{u+1}} \right) e^{2{\cal D}uv} dt^2 +
m^2(u+1)^2 e^{2{\cal D}v(2-u) - {\cal D}^2 [u^2(1-v^2) +v^2]} \left[
{{du^2}\over{u^2-1}} \right. \nonumber \\ && +
\left. {{dv^2}\over{1-v^2}} \right] + (u+1)^2(1-v^2) e^{-2{\cal D}uv}
d\varphi^2, \label{mdipole}
\end{eqnarray}
 with
\begin{eqnarray}
u&=&{1\over{2m}} \left[ \sqrt{\rho^2 + (z+m)^2} +\sqrt{\rho^2
+(z-m)^2} \right], \nonumber \\ &=&{r\over{m}}-1, \; u \geq 1,
\nonumber \\ v&= &{1\over{2m}} \left[ \sqrt{\rho^2 + (z+m)^2}
-\sqrt{\rho^2 +(z-m)^2} \right], \\ &=&\cos{\theta}, \; -1 \leq v \leq
1, \nonumber \\ \varphi& = &\varphi, \nonumber
\end{eqnarray}
where ($r,\theta,\varphi$), ($u,v,\varphi$)  and $(\rho,z,\varphi$)
 are  spherical,  prolate spherical, and   cylindrical coordinates,
 respectively. The metric (\ref{mdipole}) represents a monopolar core
 in the presence of an external dipolar field (${\cal D}$) that is
 associated to a distant shell or halo of matter. As we said before,
 this field can be seen as  the first approximation  to model an
 external concentration of matter  like halos and rings.

\subsection{Numerical method}

To solve equation (\ref{diffnlin}) in the space-time with metric
  (\ref{mdipole})  we assume: a) That fluid is stationary, i.e., the
  function $\Phi$ depends on time  only through the addition of $-at$,
  where $a$ is a constant related to the zeroth component of the
  velocity. b) That due to the axial symmetry of the metric the
  potential $\Phi$ does not depend on the variable  $\varphi$, and  c)
  That the fluid is a test fluid, i.e. the metric does not evolve and
  it is given {\em a priori}. Also we put $m=1.$

\begin{figure}
\epsfig{width=12.5cm,height=9cm, file=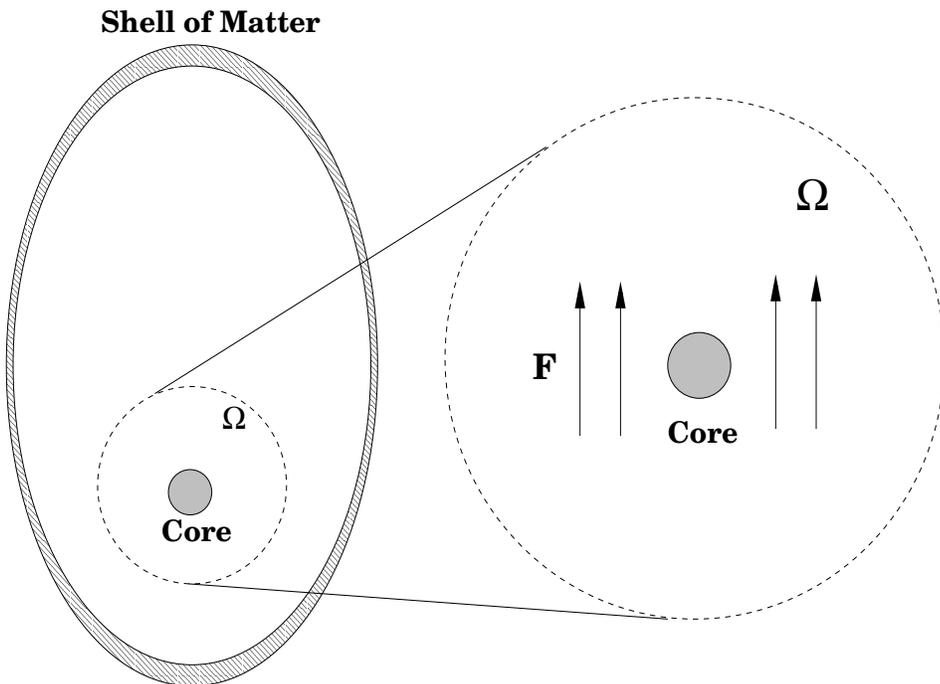}
\caption{The dipole approximation and the stationary condition leads 
to a finite region $\Omega$ of stationary flow, 
in this region a constant ``force field'' {\bf F} is present. Outside 
$\Omega$ the fluid is not  stationary.}
\label{halo}
\end{figure}

First we notice that due to the acceleration of the fluid
produced  by the dipole,   the stationary
 condition  leads to a region of instability away
from the core (black hole or hard sphere). So  we have a  region of
stability $\Omega$ that is  depicted in Fig. \ref{halo}. The size of 
this region
is very sensitive to the values of the dipolar field (${\cal D}$)
 \cite{uje:let} that accelerates the fluid.  The stationary criteria 
is checked using the fact that baryon density (\ref{density}) is a 
 positive defined quantity \cite{uje:let}.  With the
 above mentioned
 assumptions Eq. (\ref{diffnlin}) reduces to an elliptical
 differential  equation with an inner boundary conditions near the
 black hole or rigid sphere and an external boundary  condition
 (asymptotic condition).

To solve Eq. (\ref{diffnlin}) we use a   computational code  based on
  a marching method with a second order precision, five point, finite
  difference. The numerical grid is evenly spaced in $r$ and in
  $\theta$. The marching method is the {\em Stabilized Error  Vector
  Propagation} (SEVP) which is very efficient in  solving separable
  and non-separable elliptic equations. The main ingredient of the
  method is  a clever superposition of a particular solution of the
  elliptic equation with and a homogeneous solution. For a detailed
  discussion of  the method, see for instance  \cite{roa,mad}.

In the case of a rigid sphere, for the inner boundary condition,   the
usual condition of zero  normal velocity in the surface of the sphere
is employed.  In general, the fluid velocity must be equal to the
corresponding component of the velocity of the surface.  Since  usual
stars have  gaseous surfaces (not hard), this condition  does not
describe a typical flow  around a  star.  In special  astrophysical
situations, like relativistic flows around a neutron star,  this
condition can be  valid.  For a black hole we use the condition that
the fluid number density  particle   must remain finite on  the black
hole horizon \cite{pet:sha}. This leads to the numerical condition
\cite{abr:sha} that near $u=1$ (the black hole horizon),
\begin{equation}
{\partial\over{\partial r^*}} \left[ {{\Phi_{,r^*}
-\Phi_{,t}}\over{u-1}} \right] =0, \label{condtort}
\end{equation}
 where  $r^*=(u+1)+2 \ln\left( {{u-1}\over{2}} \right)$ is the
tortoise radial coordinate \cite{reg:whe}.  With the assumption that
$\Phi_{,t}=-a<0$ the condition (\ref{condtort})  can also be satisfied
in our case and it will be taken as the inner boundary condition. For 
numerical applications of this condition  see \cite{uje:let,fon:mar}.

For the outer boundary condition, with help  of the geodesic equation
for the metric (\ref{mdipole}),  we can find   a characteristic value
for the scalar field potential \cite{uje:let},
\begin{equation}
\Phi = \int \left( {{u+1}\over{u-1}} \right) e^{2 {\cal
D}v(1-u)-{{{\cal D}^2}\over{2}}[u^2(1-v^2)+v^2]} \sqrt{ \left(
{{u-1}\over{u+1}} \right) e^{2 {\cal D}uv} + k^2} \; du,
\label{integral}
\end{equation}
 where the integration is performed along the lines of constant $v$,
 usually between the surface of the sphere (or black hole
horizon) and 60  Schwarzschild radius. Later, with the results
obtained for the scalar field $\Phi$,
we check the stationary condition to see the value of the radius in which
the  system becomes non-stationary.
The constant $k$ is the fluid particle energy function.

To solve the non-linear equation (\ref{diffnlin}) we  first put  the
 non-linear term equal to  zero and calculate the linear part of the
 equation to find $\Phi$. This solution is used as an  initial guess
 for the non-linear problem. Then, from  (\ref{enthalpy}) we compute
 the fluid  enthalpy.  Finally,  with this information we  compute the
 non-linear term that is  introduced in the  non-linear equation to
 find new values of  $\Phi$. The process is repeated until the sum of
 the fractional change in the enthalpy for all interior points of the
 grid in one iteration is less than a certain error $\epsilon$.
 Usually less than 10 iterations were  required  to reach an error of
 $\epsilon \leq 10^{-6}$.  Obviously, for the linear case we need only
 one iteration. The number of iterations depends on the value of
 $\gamma$. For $\gamma \approx 1$ the nonlinear factor tends to
 infinity and the program diverge. When $\gamma=1$ the fluid is
 pressureless  (dust) and the fluid flow is geodesic, i.e., no longer
 obeys  (\ref{diffnlin}).

The code was tested  for the case $\gamma=2$  using the analytical
results for  the  steady flow  of a fluid in the presence of  either a
hard sphere \cite{sha} or  a black hole \cite{pet:sha}. In this case
the numerical solution agreed with the exact solution within an error
better than 1.5\% for the radial velocity and 1\% for the angular
velocity. Since  we are interested in qualitative aspects of the fluid
behavior  this  accuracy  is sufficient for our purposes.

 Another aspect of the fluid  dynamics is the accretion rate of matter
into a black hole.  This accretion rate can be computed from
(\ref{potflow}) and (\ref{density}), we find
\begin{equation}
\dot{N} = -\int_S nU^i \sqrt{-g} dS_i = - \int_S \kappa
h^{{{2-\gamma}\over{\gamma-1}}}\Phi_{,u} g^{uu} \sqrt{-g} dudv,
\label{accretion}
\end{equation}
where the  integration is performed in a two-surface sphere centered
on the black hole.  The exact form of  (\ref{accretion}), for
$\gamma=2$, is known  \cite{pet:sha}. It is  $\dot{N}=16 \pi M^2
n_\infty U_\infty^0$, where $n_\infty$ and  $U_\infty^0$ are  the
asymptotic density and zeroth component of the 4-velocity,
respectively.  $M$ is the black hole mass. We test our code to compute
the accretion rate with this exact solution. We found an  error  less
than 1\%.

\subsection{Numerical results}

\begin{figure}
\epsfig{width=10cm,height=10cm, file=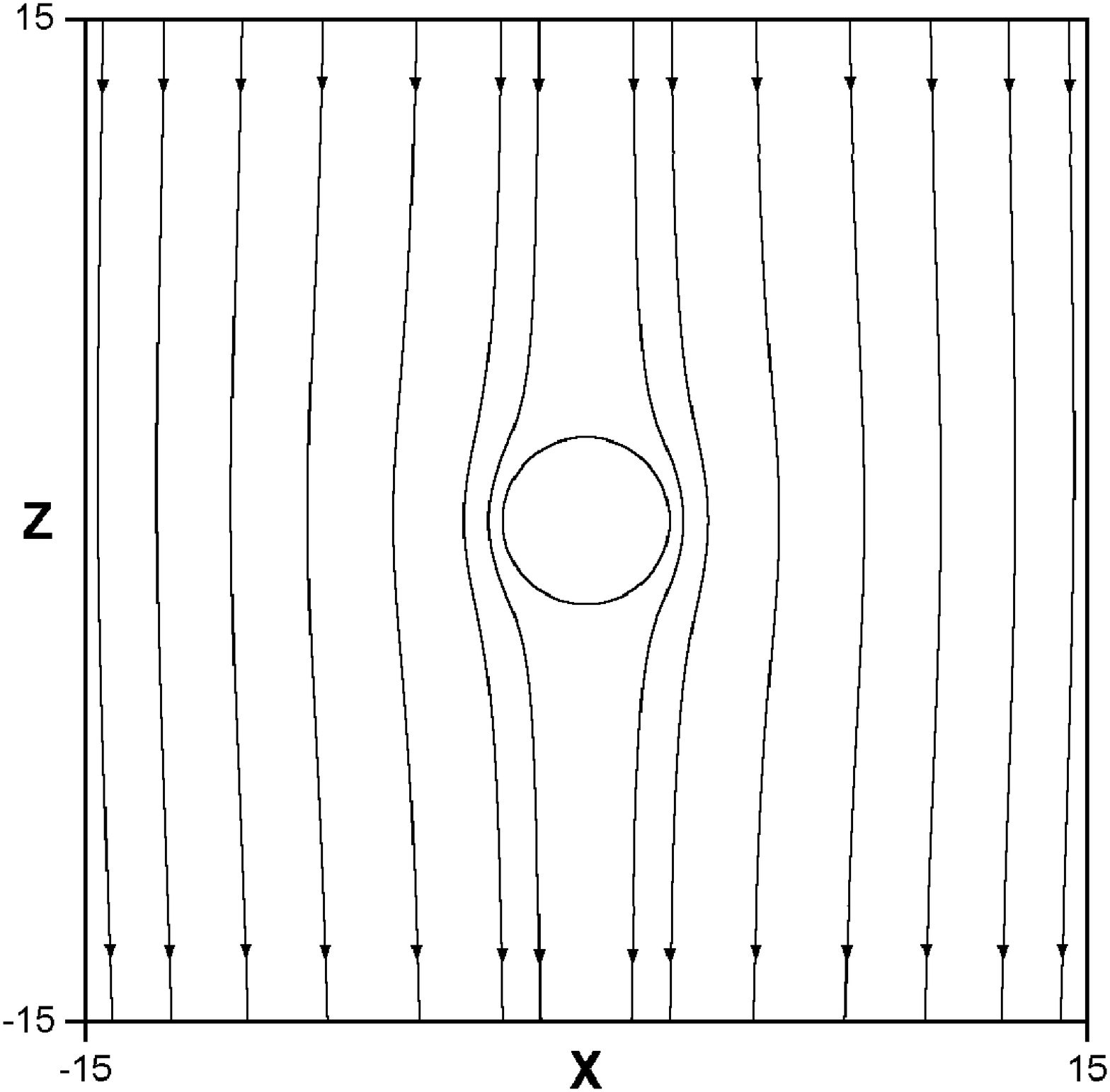}
\caption{Numerical results for the fluid streamlines with
$\gamma=1.75$  when an external dipolar field with ${\cal D}=0.001$ is
present and a rigid sphere of radius $ r = 2.5$ (Schwarzschild radius
equal 2) is placed as an obstacle. We set the values of the constant
$k=a=1.25$ . The axes are defined as $X=r \sin \theta$ and $Z=r \cos
\theta$, with $r=u+1$ and $\cos \theta = v$.}
\label{fesfcont}
\end{figure}

\begin{figure}
\epsfig{width=10cm,height=10cm, file=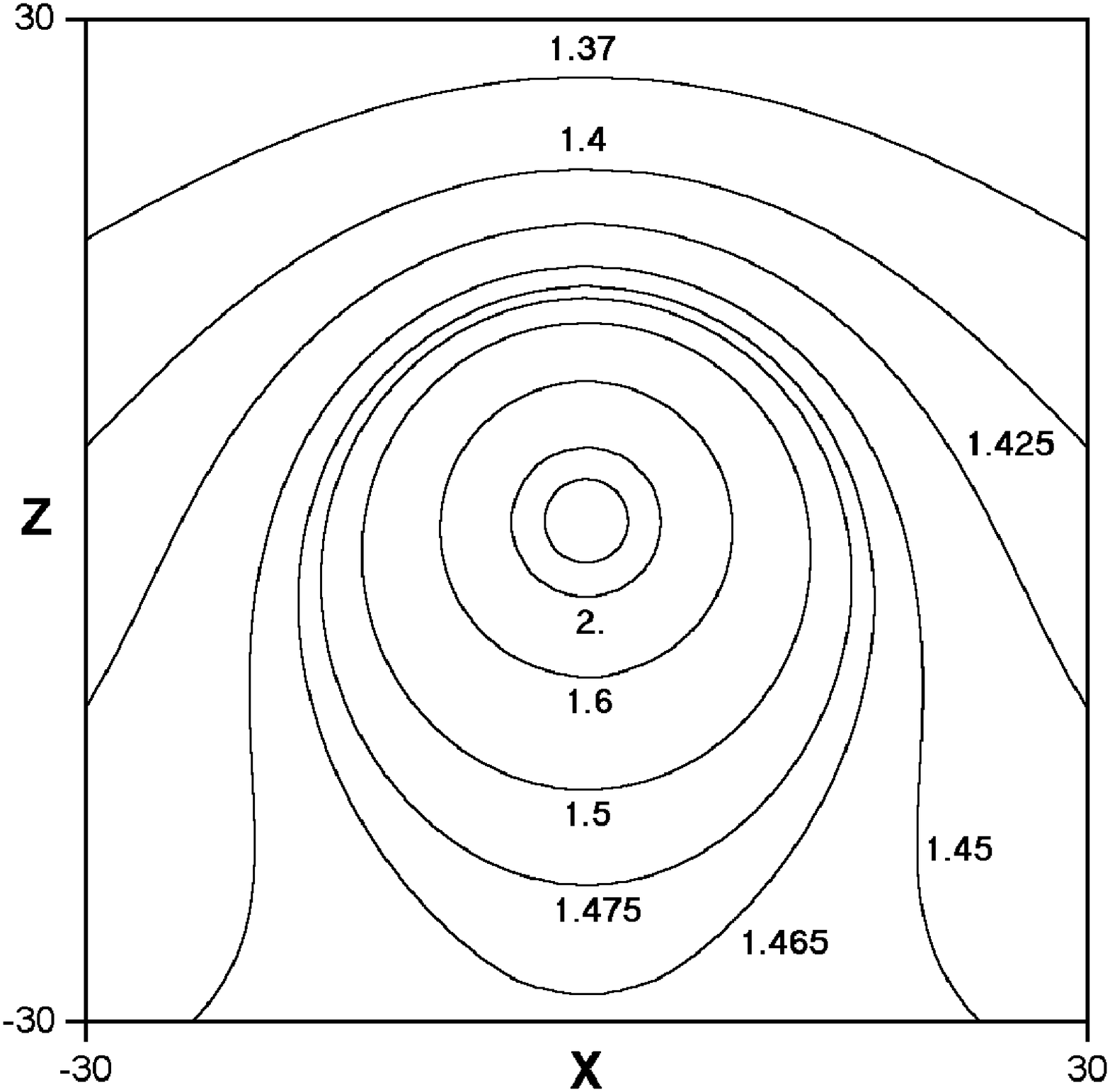}
\caption{Numerical results for the density contours ($n/\kappa$) of
the baryon number density when an external dipolar field with ${\cal
D}=0.001$ is present  and a rigid sphere of radius $r = 2.5$
(Schwarzschild radius equal 2) is placed as an obstacle. The axes and
the constants are defined as in Fig. \ref{fesfcont}.}
\label{fesfden}
\end{figure}

In Figs. \ref{fesfcont} and \ref{fesfden}, for the case of a rigid
sphere with a dipolar halo, we show the numerical results for the
streamlines  and  density contour lines of the baryon number density.
The constants are: radius of the sphere 2.5 (Schwarzschild radius 2),
dipolar strength ${\cal D}=0.001$, $\gamma=1.75$  and  $k=a=1.25$,
 the outer boundary condition was set to  60 Schwarzschild
radius. We see that the streamlines do not differ  much from the linear case
\cite{uje:let} and that the contour lines have the same qualitative
behavior as in the linear case, $(\gamma=2)$. The numerical values for
the baryon density are not the same due to the  exponent
$1/(\gamma-1)$ presented in the relation between the enthalpy and the
baryon density.  To ensure a physical result, in  solving
(\ref{diffnlin}), we  require that enthalpy   be positive in every
iteration of our code.  The baryon number density is a  positive
quantity. We can have  a negative enthalpy   solution of
$n=\kappa h^{1/(\gamma-1)}$,  but in this case  we  have a
negative baryon density.

\begin{figure}
\epsfig{width=10cm,height=10cm, file=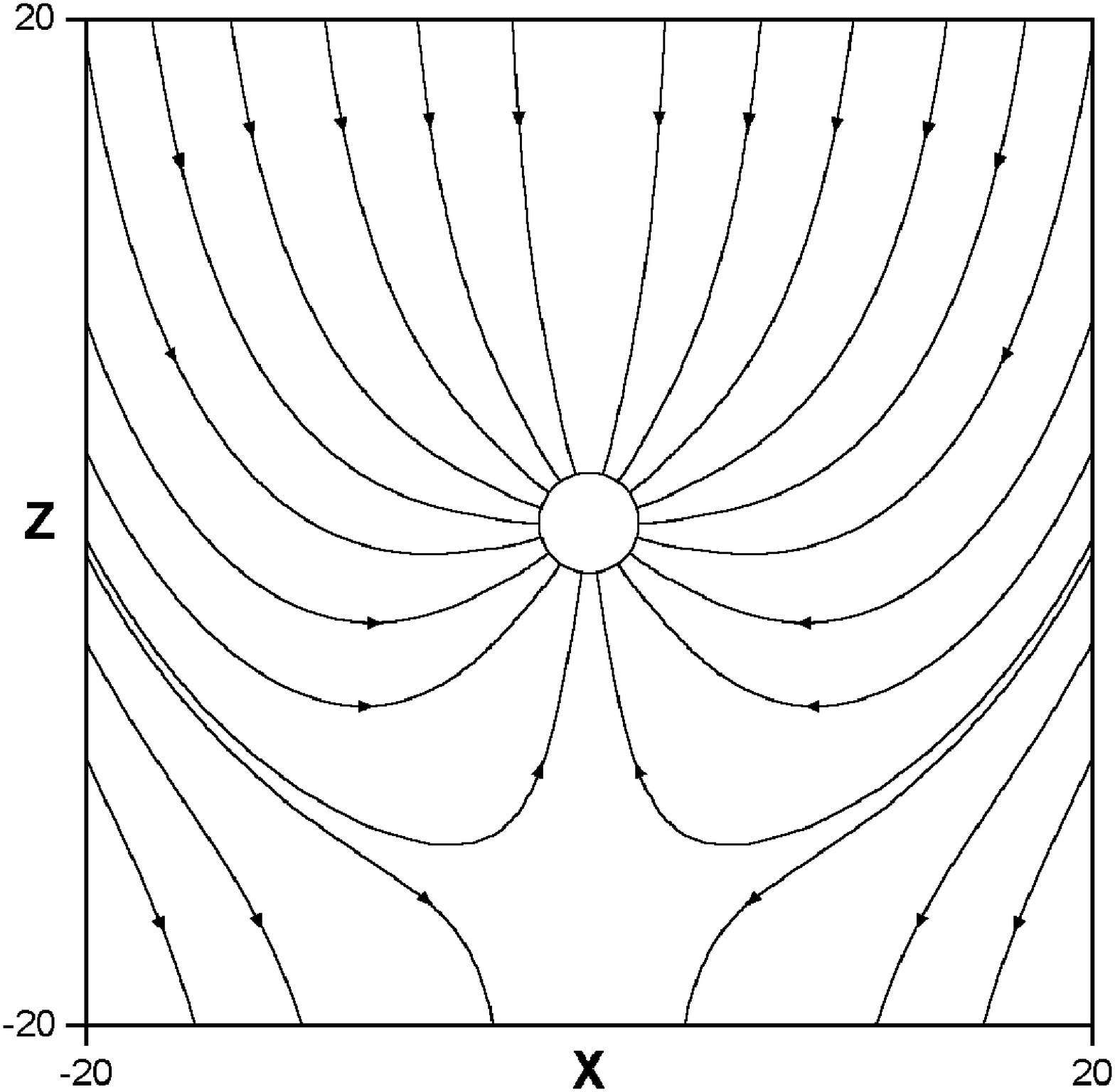}
\caption{Numerical results for the streamlines when an external
dipolar field of value ${\cal D}=0.001$ is present and a black hole is
placed as an obstacle. The black hole has radius $r=2$  (Schwarzschild
radius equal 2). The constants and the meaning of axes are the same of
Fig.  \ref{fesfcont}.}
\label{fbhcont}
\end{figure}

\begin{figure}
\epsfig{width=10cm,height=10cm, file=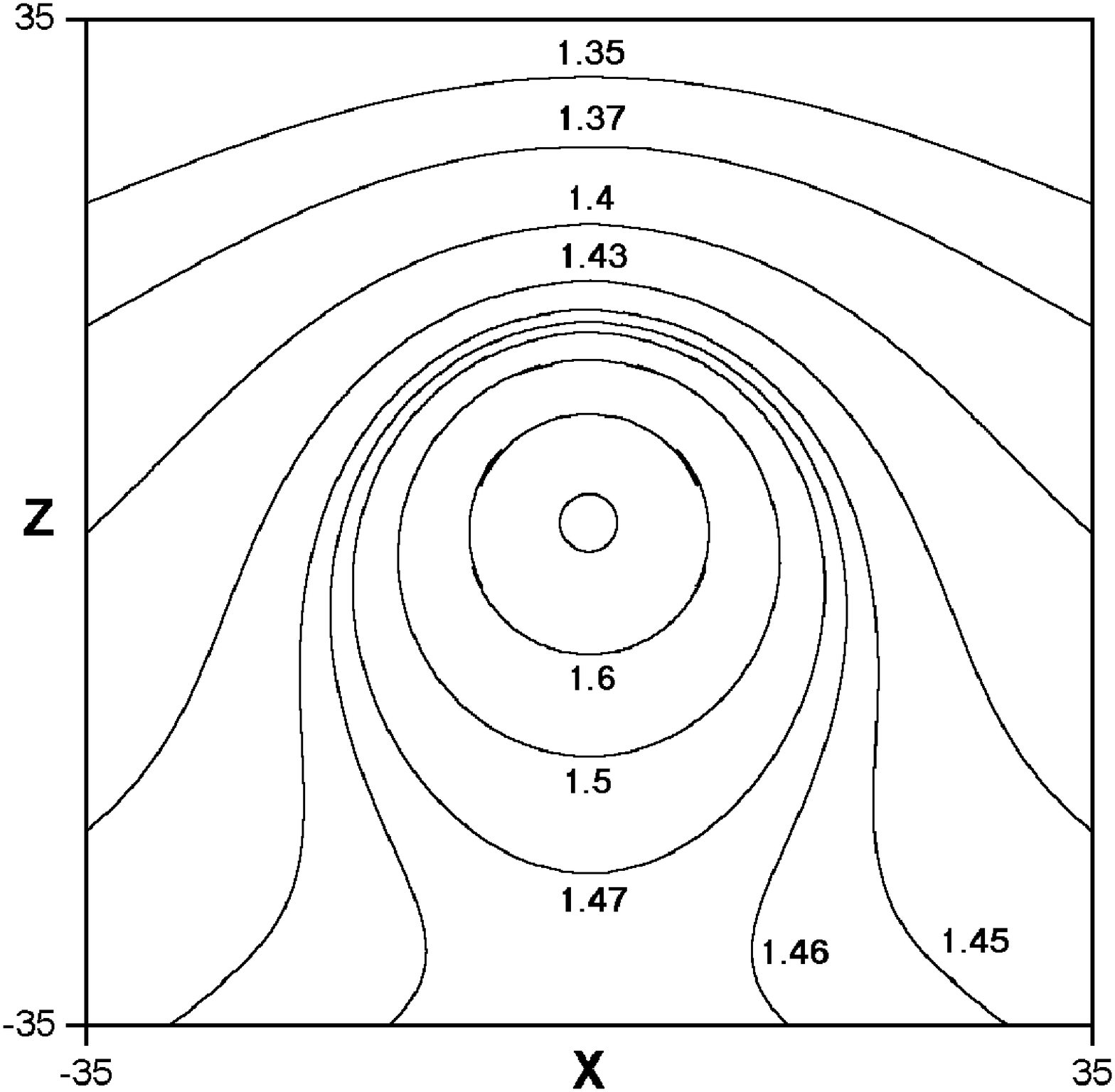}
\caption{Numerical results for the density contours ($n/\kappa$) of
the baryon number density when an external dipolar field of value
${\cal D}=0.001$ is present and a black hole is placed as an
obstacle. The black hole  has radius $r=2$ (Schwarzschild radius equal
2). The constants and the meaning of the axes are the same of
Fig. \ref{fesfcont}.}
\label{fbhden}
\end{figure}

For the same values of the constants, in Figs. \ref{fbhcont} and
\ref{fbhden} we show the numerical results for the streamlines and
density contours for the case of a black hole with a dipolar
halo. Again we see little difference compared to the linear
case. These results tell us that the contribution of the non-linear
term does not affect in a significant  way the qualitative behavior of
the fluid.

For the case of a black hole without dipole (${\cal D} = 0 $) we also
 found that, for  a  constant asymptotic velocity, that the
 streamlines and density contours do not distinguish  between   the
 linear and non linear case. The scaled accretion rate for the black hole
 ($\dot{N}/\kappa$) increases when  $\gamma$ decreases, e.g., for 
$\gamma=1.9, 1.8 ,1.7,$  the scaled accretion rate increases in $6\%, 
13\%, 24\%$, respectively, the value of the asymptotic velocity used in 
this case is $v_\infty =0.6$, the same as in \cite{pet:sha}.

For $\gamma=2$, the linear case, we find  that the value of  the scaled 
accretion rate  with or without dipolar halo differs  by $1\%$. We also 
find in this case, for the same values of Fig. \ref{fesfcont}, 
that the scaled accretion rate increases when 
$\gamma$ decreases. e.g., for $\gamma=1.9, 1.8 ,1.7,$ the scaled accretion
rate increases in $8\%, 18\%, 33\%$, respectively.  In the dipole case
the growth  of  the scaled accretion  rate when  $\gamma$ decreases is
greater than  in the case without dipole. In both cases the  growth of
the scaled accretion rate is due to the factor involving the enthalpy
[cf. Eq.  (\ref{accretion})].

As  mentioned before, for an  accelerated fluid it  is difficult to
have stationary flow. For the  cases studied along the paper we found
that for ${\cal D}\leq 0. 01$ we can have a region of  reasonable size
(a ball greater than  $45$ Schwarzschild radius) where we have laminar
flow.  For  ${\cal D}= 0.03$ the radius of this ball shrinks to $5$,
and for ${\cal D}= 0.05$ stationary flow does not exist.  In summary,
we believe that our results  are representative of a generic situation
when ${\cal D}\leq 0. 01.$

Finally, we want to point out that the variation of $\Phi$  in the
systems black hole and rigid sphere with or  without dipole for  the
linear and nonlinear cases is, in almost every interior point of
the grid, less than 5\%.

\section{ Conclusions}

We have solved numerically the streamlines and the density contours
for the baryon number density of an ideal fluid in the presence of
either a rigid sphere or a  black hole both with dipolar halo. The
fluid has barotropic equation of state, $p=(\gamma -1)$ with $ 1< 
\gamma < 2$, that represents a more realistic physical situation than 
the linear
case studied in \cite{uje:let}. We see that the nonlinear term does
not affect qualitative the form of the streamlines and density
contours when compared to the linear case, this is due to the low
variation of the scalar field $\Phi$ between the linear and nonlinear
cases. When compared with other situations studied in the literature
this fact appears to be generic.  Another interesting  result is that
the accretion rate of the cases studied increases when the constant
$\gamma$ decreases. In the presence of the dipolar field this growth
is bigger compared with the case without dipole.

The cases of a rotating black hole and rigid sphere with  halos
   modeled by  multipole moments beyond the dipole  is under active
   consideration by the authors.

\section*{Acknowledgments}

We want to thank FAPESP and CNPq for financial support.


\newpage

\newpage

\newpage

\begin{thebibliography}{99}

\bibitem{uje:let} Ujevic M and Letelier P S {\em Class. Quant. Grav.}
{\bf 18} 2917 (2001)

\bibitem{hoc:bro} Hockney R W and Brownrigg D R K {\em Mon. Not. R. 
Astron. Soc.} {\bf 167} 351 (1974)

\bibitem{hoh} Hohl F {\em Astron. J.} {\bf 83} 768 (1978)




\bibitem{abr:sha} Abrahams A M and S L Shapiro {\em Phys. Rev. D} {\bf
41} 327 (1990)

\bibitem{note1} For fluids   we  use  Landau and 
Lifshitz definitions  \cite{lan:lif}.

\bibitem{lan:lif} Landau L D and Lifshitz E M 1959 {\em Fluid
Mechanics} 2nd ed. (Oxford: Butterworth-Heinenann)

\bibitem{mon} Moncrief V 1980 {\em Astrophys. J.} {\bf 235} 1038

\bibitem{tab:tau} Tabensky R and Taub A H 1973 {\em
Commun. Math. Phys.}  {\bf 29} 61

\bibitem{vie:let} Vieira W M and Letelier P S 1999 {\em Astrophys. J.}
{\bf 513} 383

\bibitem{syn} Synge J L 1960 {\em Relativity: The General Theory}
(Amsterdan: North Holland) ch 8

\bibitem{roa} Roache P J {\em Num. Heat Transf.} {\bf 1} 1 (1978)

\bibitem{mad} Madala R V {\em Mon. Weat. Rev.} {\bf 106} 1735 (1978)

\bibitem{pet:sha} Petrich L I, Shapiro S L and Teukolsky S A 1988 {\em
Phys. Rev. Lett.} {\bf 60} 1781

\bibitem{reg:whe} Regge T and Wheeler J A 1957 {\em Phys. Rev.} {\bf
108} 1063

\bibitem{fon:mar} Font J A, Mart\'{\i} J M, Ib\'an\~ez J M and
Miralles J A {Comput. Phys. Commun.} {\bf 75} 31 (1993)

\bibitem{sha} S L Shapiro {\em Phys. Rev. D} {\bf 39} 2839 (1989)

\bibitem{note2}  We found  an oversight in equation (3.2) of
\cite{abr:sha}, in the second term (the angular term) the expression
inside the square bracket must be $(1-x^2)$ instead of
$(1-x^2)^{1/2}$. This oversight is repeated in equation (A3), (A8) and
(A10) of  the appendix concerning the finite difference schemes used
in the calculation of the particle density and the partial
differential equation. As we mentioned before the differences between
the cases  with $\gamma=2$ (linear case) and $1 < \gamma < 2$
(nonlinear case) are small.

\end{thebibliography}
\end{document}